\documentclass[preprint,showpacs,preprintnumbers,amsmath,amssymb]{revtex4}
\usepackage{graphicx}
\usepackage{dcolumn}
\usepackage{bm}

\newcommand{\be}{\begin{equation}}
\newcommand{\en}{\end{equation}}
\newcommand{\bea}{\begin{eqnarray}}
\newcommand{\ena}{\end{eqnarray}}

\begin{document}


\title{ Warm-Chaplygin  inflationary universe model  }

\author{Sergio del Campo}
 \email{sdelcamp@ucv.cl}
\affiliation{ Instituto de F\'{\i}sica, Pontificia Universidad
Cat\'{o}lica de Valpara\'{\i}so, Casilla 4059, Valpara\'{\i}so,
Chile.}
\author{Ram\'on Herrera}
\email{ramon.herrera@ucv.cl} \affiliation{ Instituto de
F\'{\i}sica, Pontificia Universidad Cat\'{o}lica de
Valpara\'{\i}so, Casilla 4059, Valpara\'{\i}so, Chile.}

\date{\today}

\begin{abstract}
 Warm inflationary universe models in the context of a  Chaplygin
 gas equation  are studied. General
 conditions required for these models to be realizable are derived and
 discussed. By using a chaotic potential we develop  models for a
   dissipation
 coefficient of the form  $\Gamma\propto\,\phi^n$, with $n=0$ or
 $n\neq 0$.
\end{abstract}

\maketitle

\section{Introduction}

It is well  know that warm inflation, as opposed to the
conventional cool inflation, presents the attractive feature that
it avoids the reheating period \cite{warm}. In these kind of
models dissipative effects are important during the inflationary
period, so that radiation production occurs concurrently together
with the inflationary expansion. If the radiation field is in a
highly excited state during inflation, and this has a strong
damping effect on the inflaton dynamics, then, it is found a
strong regimen of  warm inflation. Also, the dissipating effect
arises from a friction term which describes the processes of the
scalar field dissipating into a thermal bath via its interaction
with other fields. Warm inflation shows how thermal fluctuations
during inflation may play a dominant role in producing the initial
fluctuations  necessary for Large-Scale Structure (LSS) formation.
In these kind of models the density fluctuations arise from
thermal rather than quantum fluctuations \cite{62526}. These
fluctuations have their origin in the hot radiation and influence
the inflaton through a friction term in the equation of motion of
the inflaton scalar field \cite{1126}. Among the most attractive
features of these models, warm inflation end  at the epoch when
the universe stops inflating and "smoothly" enters in a radiation
dominated Big-Bang phase\cite{warm}. The matter components of the
universe are created by the decay of either the remaining
inflationary field or the dominant radiation field
\cite{taylorberera}.

On the other hand, the generalized Chaplygin gas    has been
proposed as an  alternative model for  describing the present
accelerating of the universe. The generalized Chaplygin gas is
described by an exotic equation of state of the form \cite{Bento}
\begin{equation}
 p_{ch} = - \frac{A}{\rho_{ch}^\beta},\label{1}
\end{equation}
where $\rho_{ch}$ and $p_{ch}$ are the energy density and pressure
of the generalized Chaplygin gas, respectively. $\beta$ is a
constant that lies in  the range $0 <\beta\leq 1$, and $A$ is a
positive constant. The original  Chaplygin gas corresponds to the
case $\beta = 1$ \cite{2}. The above equation of state leads to a
density evolution in the form \cite{Bento}

\begin{equation}
 \rho_{ch}=\left[A+\frac{B}{a^{3(1+\beta)}}\right]^{\frac{1}{1+\beta}},
 \label{2}
\end{equation}
 where $a$ is the
scale factor  and $B$ is a positive integration constant.

The Chaplygin gas emerges as a effective fluid of a generalized
d-brane in a (d+1, 1) space time, where the   action can be
written as a generalized Born-Infeld action \cite{Bento}. These
models have been extensively studied in the literature
\cite{other}. The model parameters were constrained using current
cosmological observations, such as, CMB \cite{CMB} and  supernova
of type Ia (SNIa) \cite{SIa}.

In the  model of  Chaplygin inspired inflation usually the scalar
field, which drives inflation, is the standard inflaton field,
where the energy density given by Eq.(\ref{2}), can be extrapolate
for  obtaining a successful inflationary period\cite{Ic}.
Recently, tachyon-Chaplygin inflationary universe model was
studied in Ref.\cite{SR}, and the dynamics of the early universe
and the initial conditions for inflation in a model with radiation
and a Chaplygin gas was considered  in Ref.\cite{Monerat:2007ud}.
The main goal of the present work is to investigate the possible
realization of a warm-Chaplygin inspired inflationary model, where
the universe is filled with a self-interacting scalar field and a
radiation field. We use astronomical data for constraining the
parameters appearing in this model. Specifically, the   parameters
are constrained from the WMAP observations\cite{WMAP3,WMAP}.

The outline of the paper is a follows. The next section presents a
short review of  the modified  Friedmann equation and the
warm-Chaplygin Inflationary phase. Section \ref{sectpert} deals
with the scalar and tensor perturbations.  In Section
\ref{exemple} it is presented a chaotic potential in the high
dissipation approximation. Here, we give  explicit expressions for
scalar power spectrum and tensor-scalar ratio for our models.
Finally, sect.\ref{conclu} summarizes our findings. We chose units
so that $c=\hbar=1$.

\section{The modified Friedmann equation and the Warm-Chaplygin Inflationary phase.\label{secti} }

We start by writing down  the modified Friedmann equation, by
using the FRW metric. In particular, we assume that the
gravitational dynamics  give rise to a modified Friedmann equation
of the form
\begin{equation}
H^2=\kappa\,[\rho_{ch}+\rho_\gamma]=
\kappa\left(\left[A+\rho_\phi^{(1+\beta)}\right]^{\frac{1}{1+\beta}}\;\;+\rho_\gamma\right),
\label{HC}
\end{equation}
where $\kappa=8\pi G/3=8\pi/3m_p^2$ (here $m_p$ represents the
Planck mass), $\rho_\phi=\dot{\phi}^2/2+V(\phi)$,
  $V(\phi)=V$ is
the scalar   potential and  $\rho_\gamma$ represents  the
radiation energy density.  The modification is realized from an
extrapolation of Eq.(\ref{2}), so that
\begin{equation}
 \rho_{ch}=\left[A+\rho_m^{(1+\beta)}\right]^{\frac{1}{1+\beta}}\rightarrow
 \;\left[A+\rho_\phi^{(1+\beta)}\right]^{\frac{1}{1+\beta}},
 \label{extr}
\end{equation}
where $\rho_m$ corresponds to the matter energy density \cite{Ic}.
The generalized Chaplygin gas model may be viewed as a
modification of gravity, as described in Ref.\cite{Ber}, and for
chaotic inflation, in Ref.\cite{Ic}. Different modifications of
gravity have been proposed in the last few years, and there has
been a lot of interest in the construction of early universe
scenarios in higher-dimensional models motivated by
string/M-theory \cite{Ran}. It is  well-known that these
modifications can lead to important effects in the early universe.
In the following  we will take $\beta=1$ for simplicity, which
corresponds to the usual Chaplygin gas.

 The dynamics of the
cosmological model in the warm-Chaplygin inflationary scenario is
described by the equations
 \be \ddot{\phi}+\,3H \;
\dot{\phi}+V_{,\,\phi}=-\Gamma\;\;\dot{\phi}, \label{key_01}
 \en
and \be \dot{\rho}_\gamma+4H\rho_\gamma=\Gamma\dot{\phi}^2
.\label{3}\en Here $\Gamma$ is the dissipation coefficient and it
is responsible of the decay of the scalar field into radiation
during the inflationary era. $\Gamma$ can be assumed to be a
constant or a function of the scalar field $\phi$, or the
temperature $T$, or both \cite{warm}. Here, we will take $\Gamma$
to be a function of $\phi$ only.  In the near future we hope to
study more realistic models in which $\Gamma$ not only depends on
$\phi$ but also on $T$, expression which could be derived from
first principles via Quantum Field Theory \cite{Moss,Bastero}. On
the other hand, $\Gamma$ must satisfy  $\Gamma=f(\phi)>0$ by the
Second Law of Thermodynamics. Dots mean derivatives with respect
to time, and
 $V_{, \,\phi}=\partial V(\phi)/\partial\phi$.

During the inflationary epoch the energy density associated to the
scalar field is of the order of the potential, i.e. $\rho_\phi\sim
V$, and dominates over the energy density associated to the
radiation field, i.e. $\rho_\phi>\rho_\gamma$.  Assuming the set
of slow-roll conditions, i.e. $\dot{\phi}^2 \ll V(\phi)$, and
$\ddot{\phi}\ll (3H+\Gamma)\dot{\phi}$ \cite{warm}, the Friedmann
equation (\ref{HC})  reduces  to
\begin{eqnarray}
H^2\approx\kappa\,\sqrt{A+\rho_\phi^2}\approx\kappa\,\sqrt{(A+V^2)},\label{inf2}
\end{eqnarray}
and  Eq. (\ref{key_01}) becomes
\begin{equation}
3H\left[\,1+r \;\right ] \dot{\phi}\approx-V_{,\,\phi},
\label{inf3}
\end{equation}
where $r$ is the rate defined as
\begin{equation}
 r=\frac{\Gamma}{3H }.\label{rG}
\end{equation}
For the high (weak) dissipation  regimen, we have $r\gg 1$ ($r<
1$).

We also consider that  during  warm inflation the radiation
production is quasi-stable, i.e. $\dot{\rho}_\gamma\ll 4
H\rho_\gamma$ and $ \dot{\rho}_\gamma\ll\Gamma\dot{\phi}^2$.  From
Eq.(\ref{3}) we obtained that the energy density of the radiation
field becomes
 \begin{equation}
\rho_\gamma=\frac{\Gamma\dot{\phi}^2}{4H},\label{rh}
\end{equation}
which  could be written as $\rho_\gamma= \sigma T_r^4$, where
$\sigma$ is the Stefan-Boltzmann constant and $T_r$ is the
temperature of the thermal bath. By using Eqs.(\ref{inf3}),
(\ref{rG}) and (\ref{rh}) we get
\begin{equation}
\rho_\gamma=\sigma\,T_r^4=\frac{r}{12\,\kappa\,(1+r)^2}\,\frac{V_{,\,\phi}^2}{\sqrt{A+V^2}}
.\label{rh-1}
\end{equation}
Introducing the dimensionless slow-roll parameter $\varepsilon$,
we write
\begin{equation}
\varepsilon\equiv-\frac{\dot{H}}{H^2}\simeq\frac{1}{6\kappa\,(1+r)}\,
\frac{V\,V_{,\,\phi}^2}{(A+V^2)^{3/2}},\label{ep}
\end{equation}
and the second slow-roll parameter  $\eta$
\begin{equation}
\eta\equiv-\frac{\ddot{H}}{H
\dot{H}}\simeq\,\frac{1}{3\kappa\,(1+r)\,\sqrt{A+V^2}}\,\left[V_{,\,\phi\phi}+\frac{V_{,\,\phi}^2}{V}
-\frac{3}{2}\,\frac{V\,V_{,\,\phi}^2}{(A+V^2)}\right],\label{eta}
\end{equation}
where $V_{,\,\phi\phi}=\partial^2V/\partial\phi^2$.

Note that for $r=0$ (or $\Gamma=0$), the parameters $\varepsilon$
and $\eta$ given by Eqs.(\ref{ep}) and (\ref{eta})  reduced  to
 typical expression corresponding to cool inflation where a the Chaplygin gas
is considered\cite{Ic}.

It is possible to find a relation between the energy densities
$\rho_\gamma$ and $\rho_\phi$  by using Eqs.(\ref{rh-1}) and
(\ref{ep}), so that
\begin{equation}
\rho_\gamma=\frac{r}{2(1+r)}\,\frac{(A+\rho_\phi^2)}{\rho_\phi}\,\,\varepsilon.\label{c}
\end{equation}

Warm inflation  takes place when the parameter $\varepsilon$
satisfying  the inequality $\varepsilon<1.$ This condition is
analogue to the requirement that  $\ddot{a}> 0$. The condition
given above is rewritten in terms of the energy densities
$\rho_\phi$ and $\rho_\gamma$, so that
\begin{equation}
\frac{(A+\rho_\phi^2)}{\rho_\phi}>
\frac{2(1+r)}{r}\;\rho_\gamma.\label{cond}
\end{equation}

Also, inflation ends when the universe heats up at a time when
$\varepsilon\simeq 1$, which implies
\begin{equation}
\frac{(A+\rho_\phi^2)}{\rho_\phi}\simeq
\frac{2(1+r)}{r}\;\rho_\gamma,
\end{equation}
and the number of e-folds at the end of inflation is given by
\begin{equation}
N=-3\;\kappa\,\int_{\phi_{*}}^{\phi_f}\frac{\sqrt{A+V^2}}{V_{,\,\phi}}(1+r)
d\phi\,'.\label{N}
\end{equation}

In the following, the subscripts  $*$ and $f$ are used to denote
to the epoch when the cosmological scales exit the horizon and the
end of  inflation, respectively.

\section{Perturbations\label{sectpert}}

In this section we will study the scalar and tensor perturbations
for our model. Note that in the  case of scalar perturbations the
scalar and the radiation fields are interacting. Therefore,
isocurvature (or entropy) perturbations are generated besides of
the adiabatic ones. This occurs because warm inflation can be
considered as an inflationary model with two basics fields
\cite{Jora1,Jora}. In this context dissipative effects  can
produce a variety of spectral, ranging between red and blue
\cite{62526,Jora}, and thus producing the running blue to red
spectral suggested by WMAP observations\cite{WMAP}.

As argued in Refs.\cite{warm,Liddle}, the density perturbation
could be written as
$\delta_H=\frac{2}{5}\frac{H}{\dot{\phi}}\,\delta\phi$. From
Eqs.(\ref{inf3}) and (\ref{rG}), this expression  becomes
\begin{equation}
\delta_H^2=\frac{36\,\kappa^2}{25}\left[\frac{(A+V^2)\,(1+r)^2}{V_{\,,\,\phi}^2}\right]
\,\delta\phi^2.\label{331}
\end{equation}

In the case of high dissipation, the dissipation coefficient
$\Gamma$ is much greater that the  rate expansion $H$ , i.e.
$r=\Gamma/3H\gg 1$ and following  Taylor and Berera\cite{Bere2},
we can write
\begin{equation}
(\delta\phi)^2\simeq\,\frac{k_F\,T_r\,}{2\,\pi^2},\label{del}
\end{equation}
where  the wave-number $k_F$ is defined by $k_F=\sqrt{\Gamma
H/V}=H\,\sqrt{3 r}\geq H$, and corresponds to the freeze-out scale
at which dissipation damps out to the thermally excited
fluctuations. The freeze-out wave-number $k_F$ is defined at the
point where the inequality $V_{,\,\phi\,\phi}< \Gamma H$, is
satisfied \cite{Bere2}.

From Eqs. (\ref{331}) and (\ref{del}) it follows that
\begin{equation}
\delta^2_H\approx\;\frac{18\,\sqrt{3}}{25\,\pi^2}\,\frac{T_r}{V_{\,,\,\phi}^2}\,\,
(\kappa\,r\,\sqrt{A+V^2}\,)^{5/2}\, .\label{dd}
\end{equation}

The scalar spectral index $n_s$ is given by $ n_s -1 =\frac{d
\ln\,\delta^2_H}{d \ln k}$,  where the interval in wave number is
related to the number of e-folds by the relation $d \ln k(\phi)=-d
N(\phi)$. From Eq.(\ref{dd}), we get
\begin{equation}
n_s  \approx\,
1\,-\,\left[5\widetilde{\varepsilon}-2\widetilde{\eta}-\zeta
\right],\label{ns1}
\end{equation}
where, the  parameters $\widetilde{\varepsilon}$,
$\widetilde{\eta}$ and $\zeta$, (for $r\gg 1$) are given by
\begin{equation}
\widetilde{\varepsilon}\approx\frac{1}{6\kappa\,r}\,
\frac{V\,V_{,\,\phi}^2}{(A+V^2)^{3/2}},\,\;\;\;\;\;
\,\,\widetilde{\eta}\approx\,\frac{1}{3\kappa\,r\,\sqrt{A+V^2}}\,\left[V_{,\,\phi\phi}+\frac{V_{,\,\phi}^2}{V}
-\frac{3}{2}\,\frac{V\,V_{,\,\phi}^2}{(A+V^2)}\right],
\end{equation}
and
\begin{equation}
\zeta\approx\frac{V_{,\,\phi}^2}{\kappa\,r\,\sqrt{A+V^2}}\,\left(\frac{V}{(A+V^2)}-\frac{2}{3V}\right)+
\frac{V_{,\,\phi}}
{3\kappa\sqrt{A+V^2}}\,\frac{r_{,\,\phi}}{r^2},
\end{equation}
respectively.

One of the interesting features of the observations from WMAP is
that it hints at a significant running in the scalar spectral
index $dn_s/d\ln k=\alpha_s$ \cite{WMAP}.  From Eq.(\ref{ns1}) we
obtain that the running of the scalar spectral index becomes

\begin{equation}
\alpha_s\approx\frac{2\,\widetilde{\varepsilon}\,(A+V^2)}{V\,V_{,\,\phi}}\,
[\,5\,\widetilde{\varepsilon}_{,\phi}-2\,\widetilde{\eta}_{,\phi}-\zeta_{,\phi}\,]\label{dnsdk}.
\end{equation}
In models with only scalar fluctuations the marginalized value for
the derivative of the spectral index is approximately $-0.03$ from
WMAP five year data\cite{WMAP}.

As it was mentioned in Ref.\cite{Bha} the generation of tensor
perturbations during inflation would  produce stimulated emission
in the thermal background of gravitational wave. This process
changes the power spectrum of the tensor modes by an extra
temperature dependently  factor given by $\coth(k/2T)$. The
corresponding spectrum  becomes
\begin{equation}
A^2_g=\frac{16\pi}{m_p^2}\left(\frac{H}{2\pi}\right)^2\,\coth\left[\frac{k}{2T}\right]
\simeq\frac{4\,\kappa}{\pi\,m_p^2}\,(A+V^2)^{1/2}\coth\left[\frac{k}{2T}\right],\label{ag}
\end{equation}
where the spectral index $n_g$, results to be given by $
n_g=\frac{d}{d\,\ln k}\,\ln\left[
\frac{A^2_g}{\coth[k/2T]}\right]=-2\,\varepsilon$.  Here, we have
used that  $A^2_g\propto\,k^{n_g}\,\coth[k/2T]$ as described in
Ref.\cite{Bha}.

For $r\gg1$ and from expressions (\ref{dd}) and (\ref{ag}) we may
write  the tensor-scalar ratio as
\begin{equation}
R(k)=\left.\left(\frac{A^2_g}{P_{\cal R}}\right)\right|_{\,k_*}
\simeq\left.\frac{m_p}{3\,\sqrt{8\,\pi}}
\left[\frac{V_{\,,\,\phi}^2}{T_r\,\,r^{5/2}\,(A+V^2)^{3/4}}\,\coth\left(\frac{k}{2T}\right)\right]\right|_{\,k=k_*}.
\label{Rk}\end{equation} Here, $\delta_H\equiv\,2\,P_{\cal
R}^{1/2}/5$ and  $k_*$  is referred to $k=Ha$, the value when the
universe scale  crosses the Hubble horizon  during inflation.

Combining  WMAP observations \cite{WMAP3,WMAP} with the Sloan
Digital Sky Survey (SDSS) large scale structure surveys
\cite{Teg}, it is found an upper bound for $R$ given by
$R(k_*\simeq$ 0.002 Mpc$^{-1}$)$ <0.28\, (95\% CL)$, where
$k_*\simeq$0.002 Mpc$^{-1}$ corresponds to $l=\tau_0 k\simeq 30$,
with the distance to the decoupling surface $\tau_0$= 14,400 Mpc.
The SDSS  measures galaxy distributions at red-shifts $a\sim 0.1$
and probes $k$ in the range 0.016 $h$ Mpc$^{-1}$$<k<$0.011 $h$
Mpc$^{-1}$. The recent WMAP observations results give the values
for the scalar curvature spectrum $P_{\cal
R}(k_*)\equiv\,25\delta_H^2(k_*)/4\simeq 2.3\times\,10^{-9}$ and
the scalar-tensor ratio $R(k_*)\simeq0.055<0.2$. We will make use
of these values  to set constrains on the parameters for our
model.

\section{Chaotic potential in the high dissipation approach \label{exemple}}
Let us consider  an inflaton scalar field $\phi$  with a chaotic
potential. We write for the chaotic potential as $V=m^2\phi^2/2$,
where $m$ is the mass of the scalar field. An estimation of this
parameter is given for cool-Chaplygin inflation  in Ref.\cite{Ic}
and for warm inflation in Ref.\cite{Bere2}. In the following, we
develop models for constant and variable dissipation coefficient
$\Gamma$, and  we will restrict ourselves to the high dissipation
regimen, i.e. $r\gg 1$.

\subsection{ $\Gamma =const.=\Gamma_0$
case.\label{GammaConst}}

By using the chaotic potential, we find that from Eq.(\ref{inf3})
\begin{equation}
\dot{\phi}=-\frac{m^2\,\phi}{\Gamma_0}\Longrightarrow\,\phi(t)=\phi_0\,e^{-m^2\,t/\Gamma_0},
\end{equation}
and  during the inflationary scenario the scalar field decays due
to dissipation into the radiation field. The Hubble parameter is
given by
\begin{equation}
H(t)=\sqrt{\kappa}\,\left(A+\frac{m^4}{4}\,\phi_0^4\,e^{-2m^2\,t/\Gamma_0}\right)^{1/4}.
\end{equation}
Note that in the limit $A\longrightarrow 0$ the Hubble parameter
coincide with Ref.\cite{Bere2}. The dissipation parameter $r$ in
this case is
$$
r(t)=\frac{\Gamma_0}{3\,\kappa^{1/2}}\,\left(A+\frac{m^4}{4}\,\phi_0^4\,e^{-2m^2\,t/\Gamma_0}\right)^{-1/4},
$$
where we observe that it increases when the cosmological time
increases.

By using Eq.(\ref{c}), we can relate the energy density of the
radiation field to the energy density of the inflaton field to
\begin{equation}
\rho_\gamma=\sqrt{\frac{6}{\pi}}\;\left(\frac{m^2\,m_p}
{8\,\Gamma_0}\right)\,\frac{\rho_\phi}{(A+\rho_\phi^2)^{1/4}}.\label{9}
\end{equation}
Note again that in the limit $A\longrightarrow 0$, Eq.(\ref{9})
coincides with that  corresponding to the case where the Chaplygin
gas is absent\cite{Bere2}, i.e.
$\rho_\gamma\propto\rho_\phi^{1/2}$.

From Eq.(\ref{dd}),  we obtain that the scalar power spectrum
becomes
\begin{equation}
P_{\cal R}(k)\approx\left.
\;\frac{9\;\sqrt{3}}{4\,\pi^2}\,\frac{T_r}{m^2\,V}\,
\;\left[\frac{\kappa^{1/2}\,\Gamma_0\,(A+V^2)^{1/4}}{3}\right]^{5/2}\right|_{\,k=k_*},\label{ppp}
\end{equation}
and from Eq.(\ref{Rk}) the tensor-scalar ratio is given by
\begin{equation}
R(k)\approx\;\left.\frac{m_p}{3\,\sqrt{2\pi}}\;
\left(\frac{3\,\kappa^{1/2}}{\Gamma_0}\right)^{5/2}\;\left[\frac{m^2\,V}{T_r\,(A+V^2)^{1/8}}\,
\;
\;\coth\left(\frac{k}{2T}\right)\right]\right|_{\,k=k_*}.\label{rrrr}
\end{equation}

By using the WMAP observations where $P_{\cal R}(k_*)\simeq
2.3\times 10^{-9}$, $R(k_*)=0.055$, and choosing the parameter
 $T\simeq\,T_r$, we obtained from Eqs.(\ref{ppp}) and
(\ref{rrrr}) that
\begin{equation}
V_*\simeq\frac{4\times10^{-5}\,T_r}{m^2}\,\left[\frac{m_p^2\,\Gamma_0^{2}}{\coth(k_*/2T_r)}\right]^{5/4},\label{V}
\end{equation}
and
\begin{equation}
A=\frac{10^{-21}\,m_p^8}{\coth^2(k_*/2T_r)\,}\,-V_*^2.\label{A}
\end{equation}
From Eqs.(\ref{V}) and (\ref{A}), and since $A>0$, we obtained a
lower limit for $m$ given by
\begin{equation}
m^2>10^{6}\,T_r\;\left[\frac{\Gamma_0^5}{m_p^3\,\sqrt{\coth[k_*/2T_r]}}\right]^{1/2},
\end{equation}
and also we find an upper limit for the chaotic potential  when
the universe scale crosses the Hubble horizon  during inflation,
given by
\begin{equation}
V_*<10^{-11}\,\frac{m_p^4}{\coth[k_*/2T_r]}.
\end{equation}
Now we consider the special case in which we fixe
$T\simeq\,T_r\simeq 0.24 \times 10^{16}$ GeV and
$\Gamma_0\simeq0.5\times 10^{13}$ GeV. In this special case we
obtained that the lower limit for the square mass of the scalar
field, is given by  $m^2>4\times10^{-15}\,m_p^2$, and for the
chaotic potential we obtained an upper limit given by
$V_*<10^{-14}m_p^4$.

\begin{figure}[th]
\includegraphics[width=5.0in,angle=0,clip=true]{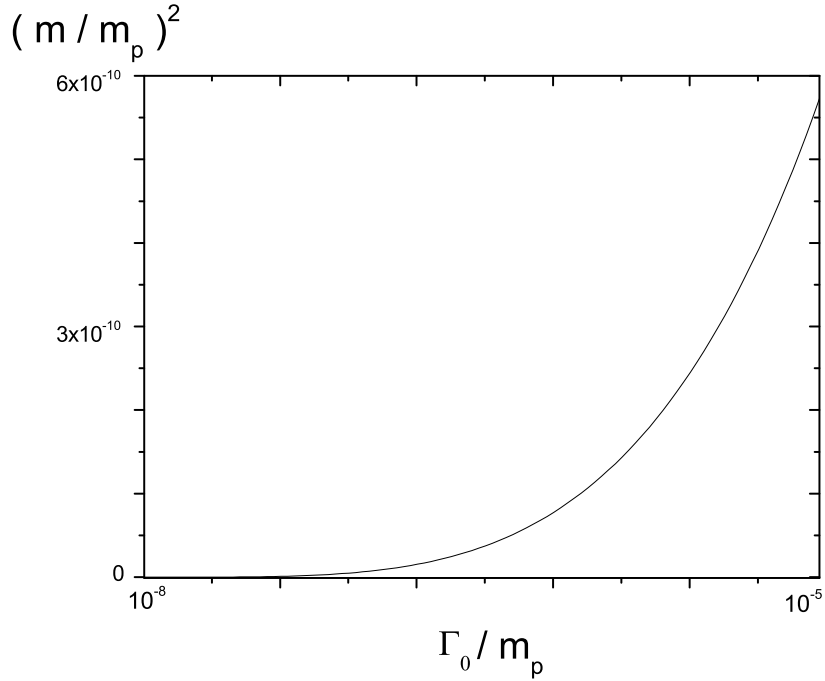}
\caption{In this figure we plot the  parameter $(m/m_p)^2$ as a
function of the dissipation coefficient $\Gamma_0/m_p$. Here, we
have taken the values $T\simeq\,T_r\simeq 0.24 \times 10^{16}$ GeV
and $k_*=0.002$ Mpc$^{-1}$.
 \label{fig11}}
\end{figure}

  In Fig.(\ref{fig11})  we have plotted the parameter
$(m/m_p)^2$ versus the dissipation coefficient $\Gamma_0/m_p$. In
doing this, we have used Eqs. (\ref{ns1}), (\ref{V}) and
(\ref{A}). Here, we have taken the values $T\simeq\,T_r\simeq 0.24
\times 10^{16}$ GeV and $k_*=0.002$ Mpc$^{-1}$.

\subsection{ $\Gamma \propto \,\phi^n$
case.\label{Gammavariable}}

In this case we consider a power-law dissipation coefficient given
by
\begin{equation}
\Gamma(\phi)=\alpha_n\,\phi^n,
\end{equation}
where $n$ is a positive integer number and $\alpha_n$ is a
positive constant. The case $n=1$ was studied in Ref.\cite{62526},
the case $n=2$ was developed in Ref. \cite{Jora1}, and the general
case for any $n$ in Ref.\cite{Lee:1999iv}.


From Eq.(\ref{inf3}) we find that
\begin{equation}
\phi(t)=\left[\phi_i^n-\frac{n\,m^2}{\alpha_n}\,t\right]^{1/n},
\end{equation}
where $\phi(t=0)=\phi_i$. The dissipation parameter, $r$, as
function of the cosmological time becomes
\begin{equation}
r(t)=\frac{\alpha_n}{3\,\kappa^{1/2}}\,\left[\phi_i^n-\frac{n\,m^2}{\alpha_n}\,t\right]\,
\left[A+\frac{m^4}{4}\,\left(\phi_i^n-\frac{n\,m^2}{\alpha_n}\,t\right)^{4/n}\right]^{-1/4},
\end{equation}
and in the high dissipation approach it is necessary to satisfy
that
$$
\frac{\alpha_n}{3\,\kappa^{1/2}}\,\left[\phi_i^n-\frac{n\,m^2}{\alpha_n}\,t\right]\gg\,\,
\left[A+\frac{m^4}{4}\,\left(\phi_i^n-\frac{n\,m^2}{\alpha_n}\,t\right)^{4/n}\right]^{1/4}.
$$

Using Eq.(\ref{c}), we can relate the energy density of the
radiation field to the energy density of the inflaton field as
\begin{equation}
\rho_\gamma=\sqrt{\frac{6}{\pi}}\;\left(\frac{m^{2+n}\,m_p}
{8\,2^{n/2}\,\alpha_n}\right)\,\frac{\rho_\phi^{1-n/2}}{(A+\rho_\phi^2)^{1/4}}.\label{91}
\end{equation}

From Eq.(\ref{dd}),  we obtain that the scalar power spectrum
becomes
\begin{equation}
P_{\cal R}(k)\approx\left.
\;\frac{9\;\sqrt{3}}{4\,\pi^2}\,\frac{T_r\,V^{\frac{5n-4}{4}}}{m^{\frac{5n+4}{2}}}\,
\;\left[\frac{2^{n/2}\,\kappa^{1/2}\,\alpha_n\,\,(A+V^2)^{1/4}}{3}\right]^{5/2}\right|_{\,k=k_*},\label{p2}
\end{equation}
and from Eq.(\ref{Rk}) the tensor-scalar ratio becomes given by
\begin{equation}
R(k)\approx\;\left.\frac{m_p}{3\,\sqrt{8\pi}}\;
\left(\frac{3\,\kappa^{1/2}}{\alpha_n}\right)^{5/2}\;\left[\frac{
m^{\frac{4+5n}{2}}\,\,(2\,V)^{\frac{4-5n}{4}}}{T_r\,(A+V^2)^{1/8}}\,
\;
\;\coth\left(\frac{k}{2T}\right)\right]\right|_{\,k=k_*}.\label{r2}
\end{equation}

By using the WMAP observations where $P_{\cal R}(k_*)\simeq
2.3\times 10^{-9}$,  $R(k_*)=0.055$, and choosing the parameter
 $T\simeq\,T_r$, we obtained from Eqs.(\ref{p2}) and
(\ref{r2}) that
\begin{equation}
m\approx
\left[\frac{3.7\times10^{-4}}{a^{1/4}\,b^{5/4}\,V_*^{\frac{4-5n}{4}}}\right]^{\frac{2}{4+5n}},\label{m2}
\end{equation}
and
\begin{equation}
A\approx\,5.9\times 10^{-11}\,\left[\frac{8.1\times
10^{-10}}{a^2\,b^2}-1.7\times 10^{10}\,V_*^2\right]\label{A2},
\end{equation}
where $a$ and $b$ are given by
$$
a=\frac{2^{\frac{5n-8}{4}}\,T_r\,\kappa^{5/4}\,\alpha_n^{5/2}}{\pi^2},\;\;\;\mbox{and}\;\;\;\;
b=\frac{m_p\,3^{3/2}}{\sqrt{\pi}\,\,2^{\frac{2+5n}{4}}}\;\frac{\kappa^{5/4}}{\alpha_n^{5/2}\;T_r}\,
\coth\left(\frac{k_*}{2\,T_r}\right),
$$
respectively.

\begin{figure}[th]
\includegraphics[width=4.0in,angle=0,clip=true]{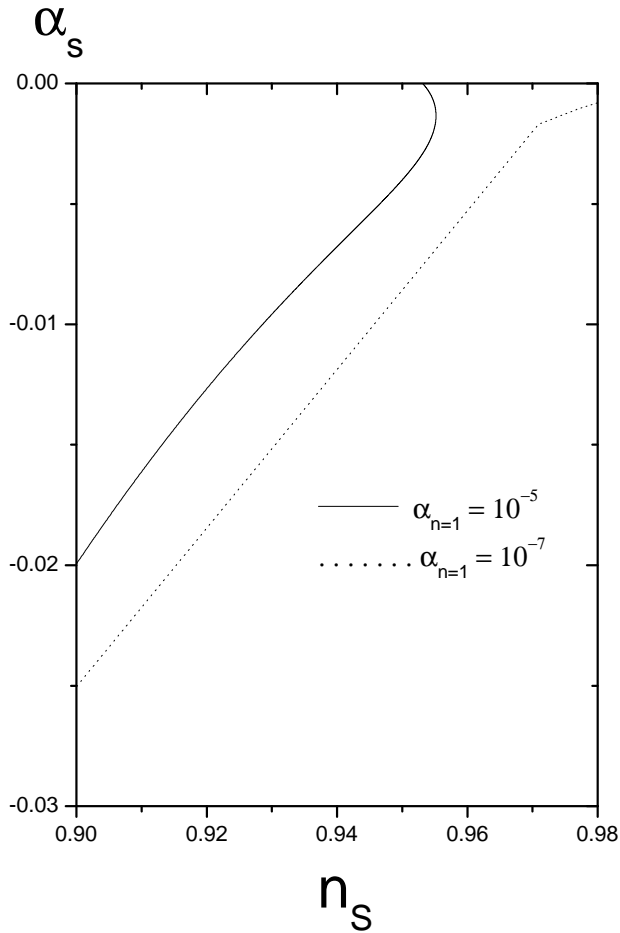}
\caption{Evolution of the  running  scalar spectral index
$\alpha_s$ versus the scalar spectrum index $n_s$ in the model
$\Gamma\propto\,\phi$, for two different values of the parameter
$\alpha_{n=1}$.
 \label{rons}}
\end{figure}

\begin{figure}[th]
\includegraphics[width=4.0in,angle=0,clip=true]{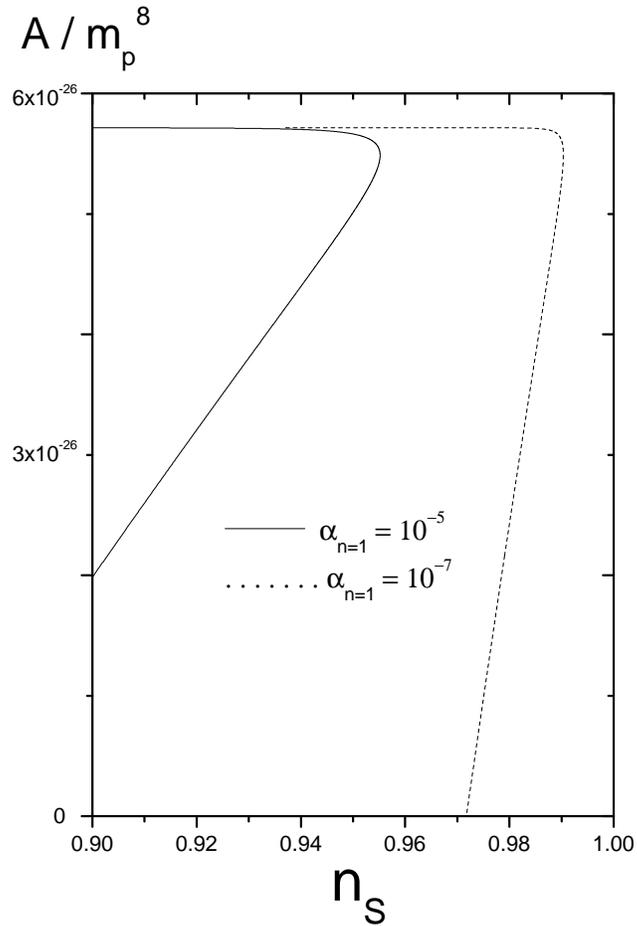}
\caption{The parameter $A$ versus the scalar spectrum index $n_s$
in the model $\Gamma\propto\,\phi$, for two different values of
the parameter $\alpha_{n=1}$.
 \label{rons2}}
\end{figure}

Using that $A>0$, then we find from Eqs. (\ref{m2}) and (\ref{A2})
a lower  limit for the mass of the inflaton $m$, independent of
the value $V_*$, and it  becomes

\begin{equation}
m\,>\,\left[\frac{a^{3-5n}\;10^{\frac{44-95n}{2}}}{b^{1+5n}}\right]^{\frac{1}{2(4+5n)}}\label{mass}.
\end{equation}
For instance, for $n=1$ we obtained from Eq.(\ref{mass}), that
$m>10^{-1}\,[T_r^2\,\alpha_{n=1}^5/\kappa^5]^{1/9}\,[m_p\,\coth(k_*/2T_r)]^{-1/3}$$\sim
7\times10^{-7}m_p$. Here, we have used that $T_r\simeq
10^{-3}m_p$, $k_*=0.002$Mpc$^{-1}$ and $\alpha_{n=1}\simeq
10^{-5}$ (see Ref.\cite{62526}).  Note that this lower limit for
$m$ decreases when  the parameter $\alpha_{n=1}$ decreases. For
$n>1$ we noted that a similar situation occurs for the lower limit
of $m$, i.e., this limit decreases when  the parameter
$\alpha_{n}$ decreases. In Fig.\ref{rons} we have plotted  the
running spectral index $\alpha_s$ versus the scalar spectrum index
$n_s$, for a dissipation coefficient
$\Gamma(\phi)=\alpha_{n=1}\,\phi$.
 In doing this, we have taken two different values for the
parameter $\alpha_{n=1}$ and choosing the parameters $T\simeq
T_r\simeq0.24\times 10^{16}$GeV and $k_*=0.002$Mpc$^{-1}$. Note
that for $\alpha_{n=1}\sim 10^{-5}-10^{-7}$ and for a given
$n_s\sim 0.95$ the values of $\alpha_s$ becomes in agrement with
that registered by the WMAP observations. From Fig. \ref{rons2} we
note that the parameter $A\sim 10^{-25} m_p^8$  becomes smaller by
one order of magnitude when it is compared with the case of
Chaplygin cool-inflation  \cite{Ic}, and smaller by two order of
magnitude when it is compared with the case of tachyon-Chaplygin
inflationary universe model \cite{SR}.

Another  interesting situation corresponds to the case when
$\Gamma=\alpha_{n=2}\,\phi^2$. Following a  similar approach to
those of the previous cases,  we obtained that $\alpha_{n=2}\sim
10^{-13}/m_p-10^{-5}/m_p$ and $A\sim10^{-26}m_p^8$ in order to be
in  agrement with  WMAP five year data\cite{WMAP}. Note that this
value of $A$ is similar to those found for the case $n=1$. Here,
again we have taken the parameters $T\simeq T_r\simeq0.24\times
10^{16}$GeV, $k_*=0.002$Mpc$^{-1}$ and $n_s\simeq 0.95$.

\begin{figure}[th]
\includegraphics[width=7.8in,angle=0,clip=true]{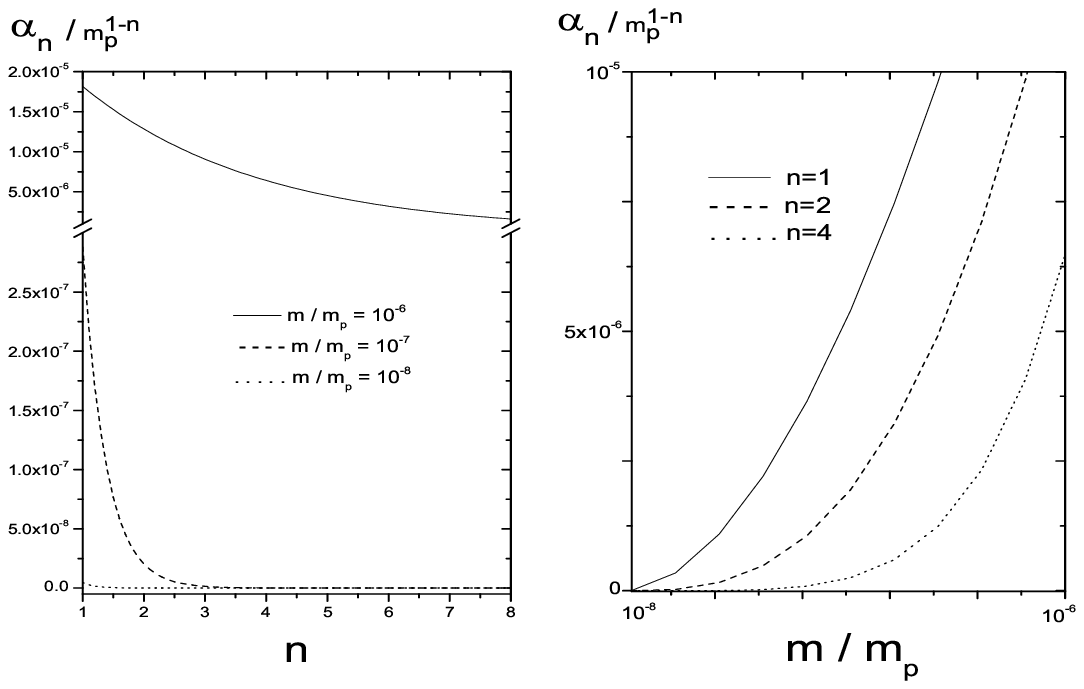}
\caption{The left panel shows the parameter $\alpha_n/m_p^{1-n}$
as a function of the parameter $n$. The right panel shows the
parameter $\alpha_n/m_p^{1-n}$ versus the mass of the inflaton
field $m/m_p$. In both cases, we have taken the parameters
$T\simeq T_r\simeq0.24\times 10^{16}$GeV and
$k_*=0.002$Mpc$^{-1}$.
 \label{fig4}}
\end{figure}

In Fig.\ref{fig4} we show the plot of the adimensional parameter
$\alpha_n/m_p^{1-n}$ in terms of the parameters $n$ on the one
hand, and $m/m_p$ in addition. In doing this, we have used Eqs.
(\ref{ns1}), (\ref{dnsdk}), (\ref{m2}) and (\ref{A2}). Here, we
have used the WMAP five year data, and  we have taken three
different values for the parameters $n$ and $m/m_p$. Also, we have
chosen $T\simeq T_r\simeq0.24\times 10^{16}$GeV and
$k_*=0.002$Mpc$^{-1}$, as above.

\section{Conclusions \label{conclu}}

In this paper we have investigated the  warm-Chaplygin
inflationary scenario. In the slow-roll approximation we have
found a general relationship between the radiation and scalar
field energy densities. This has led us to a general criterium for
 warm-Chaplygin inflation  to occur.

Our  specific models are described by  a chaotic potential and we
have consider  different  cases for   the dissipation coefficient,
$\Gamma$. In the first case, we took $\Gamma=constant=\Gamma_0 $.
Here, we have
 found  solutions for the inflaton field and the  Hubble parameter. The
 relation between  the radiation field and the inflaton field
 energy densities   presents the same
 characteristic to that corresponding to the   warm inflation case, except that
 it  depends on the extra parameter  $A$. For the case in which the
 dissipation coefficient $\Gamma$
is taken to be a function of the scalar field, i.e. $\Gamma
\propto \phi^n$, it was possible to describe an  appropriate warm
inflationary universe model for $n=1$ and $n=2$. In these cases,
we have obtained  explicit expressions for the corresponding
scalar spectrum  and the running of the scalar spectrum indices.

Finally, by using the WMAP five year data\cite{WMAP} and for the
special case where $T\simeq T_r\simeq0.24\times 10^{16}$GeV,
$k_*=0.002$Mpc$^{-1}$ and  $n_s\simeq 0.95$,  we have found new
constraints on the parameters $\alpha_n$, $A$ and $m$.

\begin{acknowledgments}
 S.d.C. was
supported by COMISION NACIONAL DE CIENCIAS Y TECNOLOGIA through
FONDECYT grant N$^0$ 1070306. Also, from UCV-DGIP N$^0$ (2008).
R.H. was supported by the ``Programa Bicentenario de Ciencia y
Tecnolog\'{\i}a" through the Grant ``Inserci\'on de Investigadores
Postdoctorales en la Academia" \mbox {N$^0$ PSD/06}.
\end{acknowledgments}


\end{document}